\documentclass[aps,prd,preprintnumbers,showpacs]{revtex4}
\usepackage{graphicx}

\begin{document}

%
%  Uncomment following two lines and one below for 2 column format.
%@twocolumnfalse\endcsname

\eprint{Nisho-1-2025}
\title{Detection of Dark Matter Axions via the Quantum Hall Effect in a Resonant Cavity}
\author{Aiichi Iwazaki}
\affiliation{Faculty of International Politics and Economics, Nishogakusha University,\\ 
6-16 3-bantyo Chiyoda Tokyo 102-8336, Japan }   
\date{July. 31, 2025}
\begin{abstract}
We propose a new method for detecting dark matter axions using a resonant cavity coupled with a quantum Hall system. When a small sample exhibiting quantum Hall effect is placed inside the cavity and the cavity is tuned to resonance, two-dimensional electrons absorb the amplified radiation, leading to a rise in the sample's temperature.
By monitoring this temperature increase, the mass $m_a$ of the axion can be inferred.
As an example, consider a GaAs sample with surface area $S=0.01\text{cm}^2$ and 
small thickness $d = 1\,\mu\mathrm{m}$ and its heat capacity $C_s$ at
temperature $T = 20\,\mathrm{mK}$. 
Because the energy flux of the incoming radiation is 
$P_{ra}\sim 5.9\times10^{-20}\text{W}\,(S/0.01\text{cm}^2)\,(g_{a\gamma\gamma}/10^{-14}\text{GeV}^{-1})^2\,(\sigma/10^7\text{eV})\,
(10^{-5}\mbox{eV}/m_a)^3(B/15\text{T})^2 
(\rho_d/0.3\rm GeV cm^{-3})$ at the resonance with electrical conductivity $\sigma$ of the cavity wall, 
the temperature increase is $P_{ra}t_{ob}/C_s \simeq 4.8\mbox{mK}(t_{ob}/1\text{s})(g_{a\gamma\gamma}/10^{-14}\text{GeV}^{-1})^2(20\mbox{mK}/T)^3 (10^{-5}\mbox{eV}/m_a)^3(\sigma/10^7\text{eV})(1\mu \text{m}/d)
(B/15\text{T})^2$ with $1\text{T}=10^4$ Gauss 
where $t_{ob}=1$s is the observation time. It must be smaller than a time constant $\tau>1$s
associated with the heat dissipation into thermal bath. Such a large time constant can be realized 
using superconducting nanowire lead and thin film pedestal supporting the sample dilution refrigerator. 
The temperature increase $\Delta T\sim 5$mK is detectable using quantum point contact thermometer.
\end{abstract}
\hspace*{0.3cm}
%\pacs{98.70.-f, 98.70.Dk, 14.80.Va, 11.27.+d \\
%Axion, Fast Radio Burst, Accretion disk}

\hspace*{1cm}

\maketitle

%\vskip2pc

\section{introduction}
Elucidating the nature of dark matter in the universe remains one of the most pressing challenges in modern physics. Among the various candidates, the axion\cite{axion1,axion2,axion3} is of particular interest as it not only serves as a compelling dark matter candidate but also offers a potential solution to the strong CP problem in quantum chromodynamics (QCD). 
Despite numerous experimental efforts ( see the references in \cite{review} ) to detect axions, they have yet to be observed.

In this paper, we propose a novel method for detecting axion dark matter using a resonant cavity coupled with the quantum Hall effect \cite{girvin}. While resonant cavities have been employed in axion detection experiments under strong magnetic fields \cite{sikivie,admx}, detecting axions with masses $m_a$ greater than $10^{-5}$eV remains challenging due to the low sensitivity caused by the small volume of the cavity. Our approach overcomes this limitation.

\vspace{0.1cm}

By tuning the cavity to an appropriate scale, microwave signals generated by axions can be resonantly amplified. When a sample exhibiting the quantum Hall effect is placed inside the cavity, the two-dimensional (2D) electrons in the quantum Hall state absorb the amplified microwave radiation, resulting in an increase in temperature. This temperature rise serves as an observable signature of axion presence and allows for the determination of the axion mass. 

\vspace{0.1cm}

For larger axion masses $m_a$, the cavity volume scales as $m_a^{-1}$ and therefore becomes smaller, leading to a corresponding reduction in the axion-induced power, which also scales as $m_a^{-2}$. As a result, direct power measurements in conventional axion haloscopes become increasingly challenging at higher axion masses.

\vspace{0.1cm}

The axion-induced power is the one absorbed in the cavity walls, 
where the microwave electric field is suppressed by a factor of $\sqrt{\sigma/m_a}$ relative to that in the cavity vacuum; here, $\sigma$ denotes the electrical conductivity of the wall. This suppression arises because the microwave is largely reflected at the conducting boundary.

In contrast, a semiconductor placed in the cavity vacuum does not strongly reflect the microwave, as it behaves as an insulator, and thus the electric field is not suppressed. Two-dimensional electrons forming an integer quantum Hall state in the semiconductor absorb the microwave radiation also without significant reflection. As a result, the absorbed power is enhanced by a factor of $\sqrt{\sigma/m_a}$ compared to that absorbed in the cavity wall. As explained below, the two dimensional electrons as we focus are in an insulator phase,
when we take Fermi energy appropriately. Thus, the significant reflection of microwaves does not arise.

This enhanced absorption leads to a measurable temperature increase in the semiconductor. Temperature changes on the order of $0.1\mathrm{mK}$ can be detected using a quantum point contact thermometer, providing a clear advantage over conventional axion haloscope techniques.

%The axion-induced power is the absorbed one in cavity wall, within which the electric field of microwave 
%is suppressed by factor $\sqrt{\sigma/m_a}$ compared with that on cavity vacuum; $\sigma$ denotes electrical conductivity of the wall.
%This is because
%the microwave is reflected by the wall. On the other hand, the microwave is not reflected by semiconductor put in the
%cavity vacuum because it is insulator. Electric field is not suppressed. Two dimensional electrons forming quantum Hall state realized 
%in the semiconductor absorbs the microwave, 
%not making it reflect. Thus, the energy power absorbed is much larger by the factor $\sqrt{\sigma/m_a}$
%than the power absorbed in the wall. It causes detectable temperature increase of the semiconductor.
%The temperature increase of the order of $0.1\text{mK}$ can be detectable with quantum point contact thermometer.
%It is an advantage over previous axion haloscopes.
We should note that the surface area of the sample is of the order of $(0.1\text{cm})^2$ or less. Only a small fraction of
induced radiations by the axion in the cavity is absorbed. 

\vspace{0.1cm}
In this paper, we consider the integer quantum Hall effect, as it allows for a rigorous treatment of microwave absorption and the heat capacity of two-dimensional electrons.

\vspace{0.1cm}
( In previous papers\cite{iwaqhe1,iwaqhe2,iwaqhe3} we have pointed out that among the experiments conducted so far on the quantum Hall effect, there exist phenomena that appear to be influenced by dark matter axions. These phenomena are thought to result from axions converting into electromagnetic waves in a strong magnetic field, which are then absorbed into the quantum Hall state. )

\vspace{0.1cm}

The temperature increase due to energy absorption depends on the heat capacity of the semiconductor sample. The total heat capacity $C$ is the sum of the semiconductor heat capacity $C_s$ and the 2D electron contribution $C_e$. By appropriately tuning the Fermi energy, the contribution from the 2D electrons $C_e$ can be made negligible compared to $C_s$; $C\simeq C_s$. Or simply by taking the volume, e.g. thickness of the semiconductor sample large, heat capacity $C_s$ can be larger than $C_e$. 
At low temperatures (e.g. $20$mK), the semiconductor heat capacity $C_s$ is dominated by phonons and follows the relation $C_s \propto T^3$, making the temperature increase highly sensitive to the temperature $T$. The lower the temperature, the greater the temperature rise for a given amount of absorbed energy. In our analysis, we assume a temperature of $T = 20$mK, which is relatively easy to achieve experimentally.

\vspace{0.1cm}
We note that the energy absorbed by electrons forming quantum Hall state is dissipated into semiconductor involving the 
two dimensional electrons. The speed of the dissipation is very rapid, for instance, within less than $10^{-6} $ second.
Thus, we may suppose 
that the absorption of the microwaves makes the temperature of semiconductor sample to increase.

\vspace{0.1cm}

We show that the microwave power $P_a$ generated by axions within the cavity is absorbed by the 2D electron system. This results in a temperature rise at a rate given by $P_{ra} / C_s$, where $P_{ra}$ is the portion of the axion-induced power actually absorbed by the 2D electrons. Without resonance, this absorbed power is negligibly small, but at resonance, it is large enough for the temperature rise to become detectable.

As a concrete example, for a GaAs sample involving the quantum Hall state, the temperature increase rate at resonance is approximately

\begin{equation}
\frac{P_{ra}}{C_s} \simeq \frac{0.72\,\mathrm{mK}}{\mathrm{s}} \times g_{\gamma}^2\left( \frac{20\,\mathrm{mK}}{T} \right)^3 
\Big(\frac{10^{-5}\mbox{eV}}{m_a}\Big)\left( \frac{B_t}{1.5 \times 10^5\,\mathrm{Gauss}} \right)^2 \left( \frac{1\,\mu\mathrm{m}}{d} \right)\Big(\frac{\rho_d}{0.3\rm GeV cm^{-3}}\Big),
\end{equation}
with the density $\rho_d$ of the dark matter axion, 
where $d$ is the sample thickness. $g_{\gamma}$ denotes QCD axion model; $g_{\gamma}(\rm KSVZ)\simeq -0.96$ and 
$g_{\gamma}(\rm DFSZ)\simeq 0.37$.
The 2D electron layer is tilted with respect to the magnetic field $\vec{B}_t$, with an angle $\theta = \pi/6$ between the layer and the plane perpendicular to $\vec{B}_t$ (see Fig.\ref{b}). This tilting makes possible absorption of the axion-induced microwave radiation.

It is important to note that this rate of temperature rise is independent of the surface area $S$ of the sample, since both the absorbed power $P_{ra} \propto S$ and the heat capacity $C_s \propto S$. The form of semiconductor sample is supposed to be rectangular with surface area $S$ and
thickness $d$.
It apparently seems that the temperature is larger as the observation time $t_{ob}$ is longer, $\Delta T=P_{ra}/C\times t_{ob}$. 

\vspace{0.1cm}
But, the absorbed energy dissipates into the thermal bath and the actual observed temperature increase $\Delta T$ depends on the time constant $\tau$ of the dissipation. The energy dissipation arises through small lines for measurement of electric current and pedestal
supporting the sample.  
However, the energy dissipation is made very slow by using weakly coupled lines and small pedestal.
For instance, we use superconducting nanowire leads and thin film support. The use of these materials make
the time constant $\tau>1$s. Then, the actual temperature increase is determined by the
the observation time $t_{ob}$ as far as $t_{ob}<\tau$.  
Ideally, weakly coupled lead wires and a pedestal with a small contact area for supporting the sample are preferred to have a large time constant $\tau>1$s. The temperature increase is

\begin{equation}
\Delta T \simeq 0.72\,\mathrm{mK} \times \left( \frac{t_{ob}}{1\,\mathrm{s}} \right)g_{\gamma}^2
\left( \frac{20\,\mathrm{mK}}{T} \right)^3\Big(\frac{10^{-5}\mbox{eV}}{m_a}\Big)\Big(\frac{B_t}{1.5\times 10^5\mbox{Gauss}}\Big)^2\left( \frac{1\,\mu\mathrm{m}}{d} \right)
\Big(\frac{\rho_d}{0.3\rm GeV cm^{-3}}\Big) 
\end{equation}
for QCD axion or for general coupling constant $g_{a\gamma\gamma}$, 

\begin{equation}
\Delta T
\simeq 5\text{mK}\left(\frac{t_{ob}}{1\,\text{s}}\right)\,\Big(\frac{g_{a\gamma\gamma}}{10^{-14}\text{GeV}^{-1}}\Big)^2
\left(\frac{20\text{mK}}{T}\right)^3\left(\frac{10^{-5}\,\text{eV}}{m_a}\right)^3
\left(\frac{B_t}{1.5\times10^5\,\text{Gauss}}\right)^2\left(\frac{1\,\mu\text{m}}{d}\right)\left(\frac{\rho_d}{0.3\,\text{GeV}\rm cm^{-3}}\right) .
\end{equation}

%It would be feasible to fabricate the sample with small lead wires and pedestal with small contact, in order to have
%time constant $\tau$ of the energy dissipation being sufficiently longer than the observation time, i.e. $1\,$s.
In order to measure the small temperature increase of the order of $0.1$mK,
it is favorable to use a thermometer based on quantum point contact, so called  QPC. 
The thermal conductance of the thermometer to thermal bath can be made also small.  

\vspace{0.1cm}

We focus on axion detection around a mass of 
$10^{-5}$eV. 
However, as indicated by these equations, 
axions with smaller masses $\sim 10^{-6}$eV are expected to produce larger temperature increases and are therefore easier to detect.

\vspace{0.1cm}

We begin by reviewing the fundamental properties of the quantum Hall state in 2D electron systems.
We then demonstrate how microwave radiation, generated by axions in the presence of an external magnetic field $B_t$, is absorbed by the 2D electrons.
Our analysis shows that the entire radiation flux entering the sample is fully absorbed in the quantum Hall state.
Through out the paper,  the temperature is of the order of $10$mK so that 
chemical potential is nearly equal to the Fermi energy $E_f$.
We use physical units, Boltzmann constant $k_B=1$, light velocity $c=1$ and $\hbar=1$.

\section{integer quantum hall effect}
\label{2}
First, we explain integer quantum Hall effect\cite{girvin}. 
Two dimensional free electrons under magnetic field $B$ form Landau levels
with their energies $E_n=\omega_c( n+1/2)$ where cyclotron frequency $\omega_c=eB/m^{\ast}\sim 10^{-2}(B/10^5\rm Gauss)$eV; 
effective mass $m^{\ast}$ of electron
in semiconductor. Generally $m^{\ast}$ is much smaller than real electron mass $m_e\simeq 0.51$MeV. ( For instance,
$m^{\ast}=0.067m_e$ in GaAs. Hereafter, we use physical parameters of GaAs. )  
 The 2D electrons are trapped in a quantum well with depth less than $10^{-2}\mu$m
between for instance, GaAs and AlGaAs. So, 
the motion of the electrons in the direction perpendicular to the well 
is forbidden in temperature, e.g. $1$K, while the motion in the direction parallel to the well is
energetically allowed.
Each Landau level possesses many degenerate states;  the number density of degenerate state per unit area
is given by the degeneracy $eB/2\pi$. Therefore,
density of states is a delta function $\propto \delta (E-E_n)$.
Their wave functions are represented using Hermite polynomials $H_n(x)$,

\begin{equation}
\Phi(\vec{x},n,k)=C_{n}H_n\Big(\frac{x-l_B^2k}{l_B}\Big)\exp\Big(-\frac{(x-l_B^2k)^2}{2l_B^2}\Big)\exp(-iky)
\end{equation} 
with normalization constant $C_{n}=(2^{n}n!\sqrt{\pi}l_BL_y)^{-1/2}$,
where $L_y$ is the length of the Hall bar in the direction $y$. 
$l_B=\sqrt{1/eB}$ is magnetic length; $l_B\simeq 8.1\times 10^{-7}\rm cm \sqrt{10^5\rm Gauss/B}$.
Degenerate states in each Landau level are specified by $k$. 
( The coordinate of the two dimensional space is $x$ and $y$ and $\vec{B}=(0,0,B)$.)

\vspace{0.1cm}
Including the spin contribution, 
each Landau level is split to two states with energies $E_{n\pm}=\omega_c(n+1/2)\pm g\mu_B B$;
$\pm$ correspond to spin parallel and anti-parallel to the magnetic field $B$, respectively. Namely,
Zeeman energy $\pm g\mu_B B$ with $g\simeq 0.44$ and Bohr magneton $\mu_B=e/2m_e$.
Typically, it is of the order of $10^{-4}(B/10^5\rm Gauss)$eV.  
Obviously Zeeman energy is smaller than the cyclotron energy in semiconductor.
The degeneracy of each Landau level with definite spin is given by $eB/2\pi$.

\section{localization of two dimensional electrons}
\label{3}

In general, real samples contain a disorder potential $V$, which lifts the degeneracy of energy levels. It is well established that, in the presence of such disorder, all 2D electron states become localized \cite{localization}. However, when a strong external magnetic field is applied, a portion of these electronic states remain delocalized. This partial delocalization is essential for the emergence of the quantum Hall effect.

\begin{figure}[htp]
\centering
\includegraphics[width=0.6\hsize]{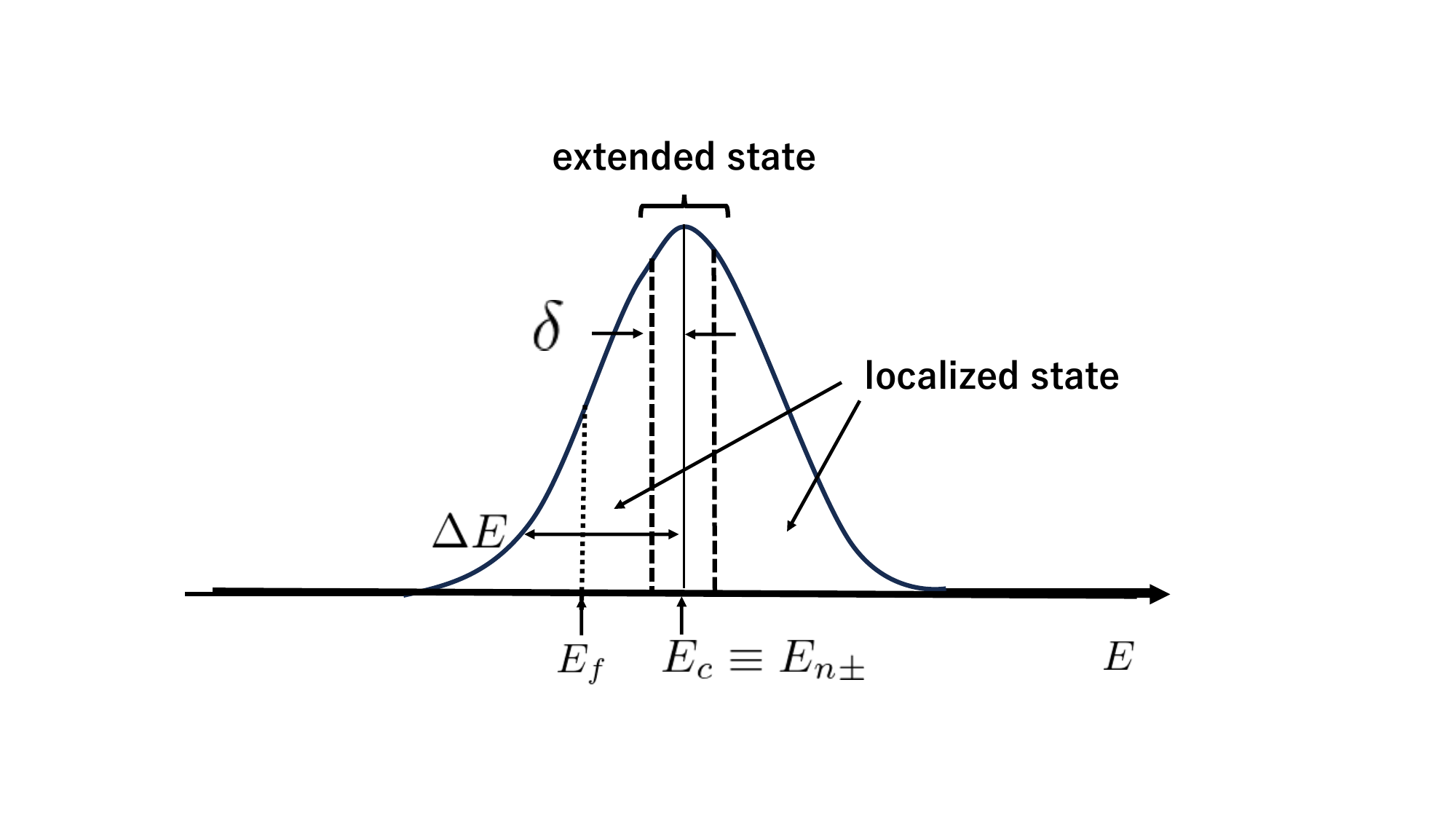}
\caption{Density of state $\rho(E)$ in finite sample. There is a region in which states are extended over the sample, while states in
other region are localized.
Dotted line represents Fermi energy $E_f$}
\label{a}
\end{figure}

Taking the disorder potential into account, the density of states ( DOS ) acquires a finite width. Numerical simulations show that the DOS, denoted $\rho(E)$, has a finite spread $\Delta E$ centered around energy levels $E_{n\pm}$, as schematically illustrated in Fig.(\ref{a}). For example, the DOS may take the form

\begin{equation}
\rho(E) \propto \exp\left( -\frac{(E - E_{n\pm})^2}{(\Delta E)^2} \right),
\end{equation}
with $E_{n\pm} \gg \Delta E$ \cite{andouemura}. This form will be used in our later discussions. The disorder potential is assumed to contain equal amounts of attractive and repulsive contributions, resulting in a symmetric DOS profile centered at $E_{n\pm}$, as shown in the figure.

The width $\Delta E$ is assumed to be much smaller than the cyclotron energy $\omega_c = eB/m^{\ast}$ under a strong magnetic field $B$. This condition is satisfied when the disorder potential energy is significantly smaller than the cyclotron energy, i.e., $V \ll \omega_c \simeq 1.8 \times 10^{-2} (B/10^5~\text{Gauss})~\text{eV}$.

\vspace{0.1cm}

Electrons occupy states with energies lower than the Fermi energy $E_f$, while states with energies higher than $E_f$ remain unoccupied. The system is characterized by the filling factor $\nu \equiv 2\pi\rho_e / eB$, where $\rho_e$ is the electron density. When $\nu < 1$, only a portion of the lowest Landau level is filled. In contrast, when $\nu > 1$, the lowest Landau level is fully occupied, and electrons begin to populate the second Landau level. Taking spin into account, all electrons still reside in the lowest Landau level as long as $\nu < 2$.

\vspace{0.1cm}

In Fig.(\ref{a}), we also indicate the energy range of extended states $E_\beta$, defined by $E_c - \delta \leq E_\beta \leq E_c + \delta$, where $E_c \equiv E_{n\pm}$. This region can be determined using a scaling law \cite{aoki1,aoki2}, which states that the localization (or coherence) length $\xi(E)$ diverges as

\begin{equation}
\xi(E) \propto |E - E_c|^{-2.4}
\quad \text{as } E \to E_c.
\end{equation}

Thus, the critical energies $E = E_c \pm \delta$ correspond to the condition where the coherence length $\xi(E)$ becomes comparable to the size of the sample. The coherence length $\xi(E)$ characterizes the spatial extent of a quantum state with energy $E$, and hence, within the interval $[E_c - \delta, E_c + \delta]$, the states are effectively extended and contribute to electrical conduction.
In infinitely large system, the interval vanishes, i.e. $\delta=0$. Only a state with energy $E_c$ is extended.

\vspace{0.1cm}

When the Fermi energy $E_f$ lies below $E_c - \delta$, electrons occupy only localized states and cannot carry current; the system is in an insulating phase. In this regime, the longitudinal conductivity $\sigma_{xx} = 0$, and the transverse conductivity $\sigma_{xy}$, or equivalently, the Hall resistance $\rho_{xy}=1/\sigma_{xy}$, exhibits quantized plateaus as a function of magnetic field. These plateaus have quantized values, such as
$\rho_{xy}=\Big(\frac{2\pi}{e^2}\Big)\Big(\frac{1}{n}\Big)\,\, \mbox{with} \,\,n\,\,\mbox{positive integer}  ,
$
arising from the topological nature of the quantum Hall effect \cite{topology1,topology2}.

On the other hand, when $E_c - \delta < E_f < E_c + \delta$, the system undergoes a plateau-to-plateau transition. In this regime, some electrons occupy extended states and can conduct electricity; the system behaves as a metal with $\sigma_{xx}\neq 0$. 
As a result, the conductivity $\sigma_{xy}$ and Hall resistance $\rho_{xy}$ change continuously with the magnetic field. It should be noted that the Fermi energy $E_f$ itself varies with the magnetic field $B$.

\vspace{0.1cm}

( In addition to insulating and metallic phases, quantum Hall systems can also exhibit a superconducting-like phase \cite{joseph1,joseph2,joseph3,iwazaki} when two parallel quantum Hall layers are placed in close proximity. Josephson-like effects between such bilayer systems have indeed been experimentally observed \cite{joseph3}. )

\vspace{0.1cm}

In this paper, we focus solely on transitions between localized states induced by axion-generated radiations. That is, the Fermi energy is taken to satisfy $E_f < E_c - \delta - m_a$, where $m_a$ is the axion mass. Under these conditions, the longitudinal conductivity $\sigma_{xx}$ remains zero even in the presence of axion-induced radiation. Consequently, the system remains in the insulating phase
so that microwaves are not reflected by the two dimensional electrons.

\section{axion dark matter}
\label{5}
In order to discuss the absorption of the axion induced radiation in quantum Hall states, we explain
electromagnetic coupling with axion\cite{axion1,axion2,axion3} and smallness of the coupling.  
The coupling $g_{a\gamma\gamma}$ between axion $a(t,\vec{x})$ and electromagnetic field is 
given by
$L_{a\gamma\gamma}=g_{a\gamma\gamma}a(t,\vec{x})\vec{E}\cdot\vec{B}$
with electric $\vec{E}$ and magnetic $\vec{B}$ fields.
The coupling $g_{a\gamma\gamma}$ is rewritten such that  $g_{a\gamma\gamma}=g_{\gamma}\alpha/f_a\pi$,
where fine structure constant $\alpha\simeq 1/137$ and axion decay constant $f_a$
satisfying the relation $m_af_a\simeq 6\times 10^{-6}\rm eV\times 10^{12}$GeV in the QCD axion.
The parameter $g_{\gamma}$ depends on the axion model, i.e. 
$g_{\gamma}(\rm DFSZ)\simeq 0.37$ for DFSZ model\cite{dfsz,dfsz1} and $g_{\gamma}(\rm KSVZ)\simeq -0.96$ for KSVZ model\cite{ksvz,ksvz1}.
Hereafter we mainly consider QCD axion model, although our results are also presented
with general $g_{a\gamma\gamma}$ independent of $m_a$. 

\vspace{0.1cm}
The coupling is much small for the axion dark matter which is described by $a(t,\vec{x})\simeq a_0\cos(m_at)$. 
( The momentum $k$ of the axion dark matter is of the order of $10^{-3}m_a$ near the Earth. 
We may neglect the momentum. )
Supposing that the local energy density $\rho_d$ of the dark matter is composed of the axion,
$\rho_d=m_a^2\overline{a(t,\vec{x})^2}=m_a^2a_0^2/2\sim 0.3\rm GeV/cm^3$ ( $\overline{Q}$ denotes
time average of the quantity $Q$. ) 
Then, we find the coupling $g_{a\gamma\gamma}a(t,\vec{x})\sim 10^{-21}$ is very small 
and it is independent of the axion mass in the QCD axion model. 

\vspace{0.1cm}
Owing to the coupling, we have oscillating electric field $E_a=g_{a\gamma\gamma}a(t)B$ in the presence of 
the axion under the magnetic field $B$. 
The electric field makes electrons in metal oscillate so that oscillating electric current in the metal
produce electromagnetic radiations. This is a mechanism of the production of electromagnetic radiations 
in resonant cavity under strong magnetic field.
 The radiation is represented by the gauge field $\vec{A}_a=\vec{A}_0\exp(-i\omega t+i\vec{p}\cdot\vec{x})\simeq \vec{A}_0\exp(-im_a t)$
whose amplitude $|\vec{A}_0|\sim g_{a\gamma\gamma}a_0B/m_a$.
When we appropriately tune the size of the cavity,
the radiations are amplified\cite{sikivie, iwazaki01} with the resonance.
Because we have cosmological constraints\cite{Wil,Wil1,Wil2},$m_a=10^{-6}\rm eV\sim 10^{-3}eV$ 
in QCD axion model,
the electromagnetic radiations are microwaves. 
Hereafter we address the microwaves in the discussion.

\vspace{0.1cm}

To detect the amplified radiation, we utilize a quantum Hall system placed inside a resonant cavity.
When a semiconductor sample exhibiting the quantum Hall effect absorbs axion induced microwaves
amplified by resonance, 
its temperature increases. The surface area of the sample is of the order of $(0.1\text{cm})^2$ or less so that only a fraction of
induced radiations by the axion in the cavity is absorbed. Putting such a small sample in the resonant cavity,
we can determine the mass of axion with the detection of the 
temperature increase. 
This provides a new method for axion detection.

In the following, we calculate the absorption energy from the axion induced microwaves 
and estimate the resulting temperature increase of a sample exhibiting the quantum Hall effect.

\section{absorption of axion microwaves in quantum Hall state}
\label{6}

\begin{figure}[htp]
\centering
\includegraphics[width=0.6\hsize]{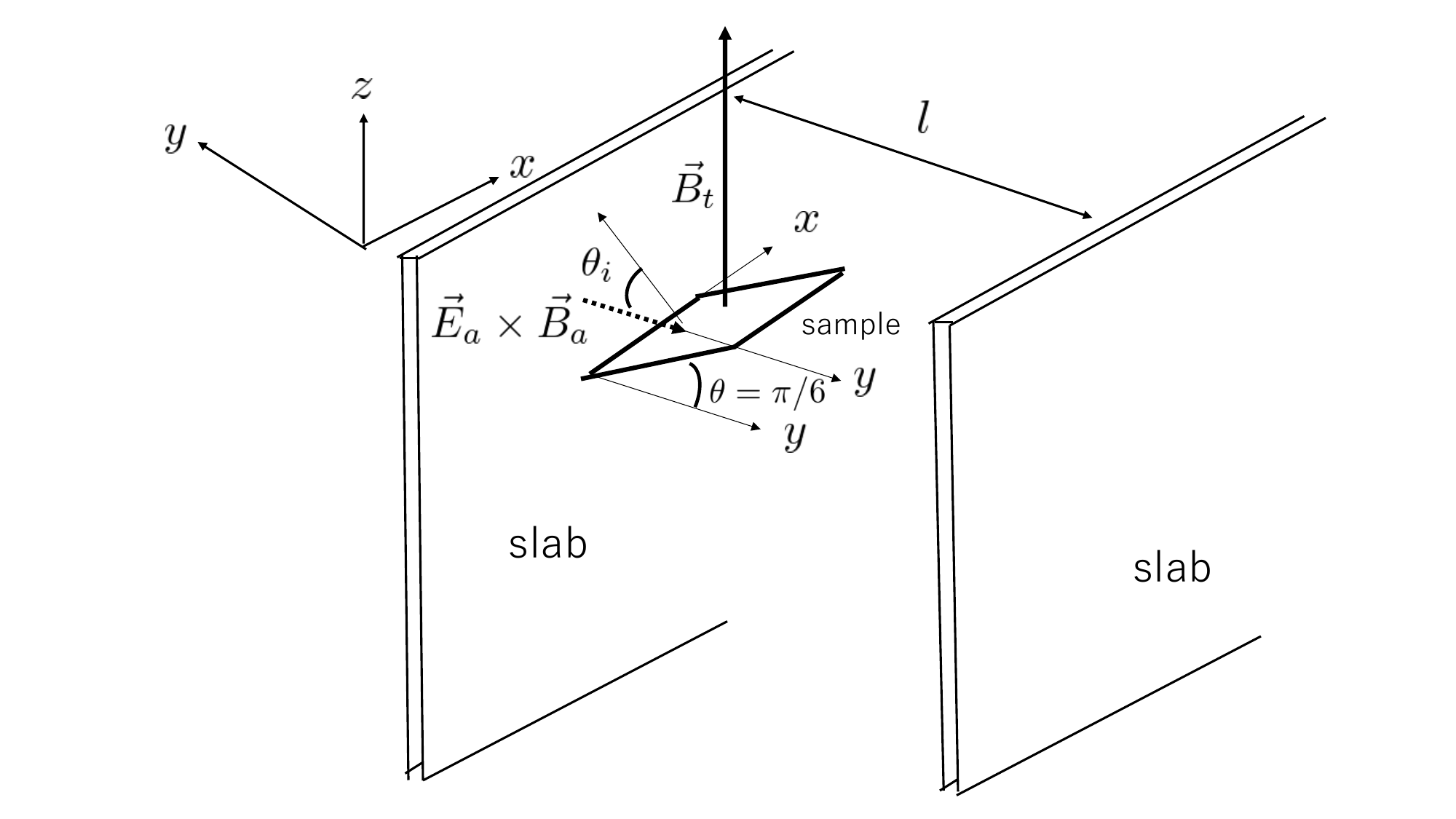}
\caption{ Slabs parallel to each other separated with distance $l$. Sample tilted with $\theta=\pi/6$ to $y$ axis
and parallel to $x$ axis. }
\label{b}
\end{figure}

To make our discussion more concrete, we consider a resonant cavity with parallel slabs, as shown in Fig.(\ref{b}), following our previous work \cite{iwazaki01}.
Let the distance between the two parallel, facing slabs be $l$, and assume that an external magnetic field $\vec{B}_t = (0, 0, B_t)$ is applied parallel to the slabs.
By adjusting the distance $l$, the resonance condition can be satisfied.
We put a semiconductor sample involving two dimensional electrons in quantum Hall state. The sample in the cavity is very small
compared with the length $l$. The surface area is of the order of $(0.1\text{cm})^2$ or less. 

Microwaves with electric field $\vec{E}_a = (0, 0, E_a)$ and magnetic field $\vec{B}_a = (B_a, 0, 0)$ are generated in the presence of the external field $\vec{B}_t$.
The resulting energy flux is $\vec{E}_a \times \vec{B}_a = (0, E_a B_a, 0)$, indicating propagation in the $y$-direction.
The wave is standing wave between the slabs.
\vspace{0.1cm}
It should be noted that the fields $E_a$ and $B_a$ vary with the spatial coordinate $y$, assuming the slabs are located at $y = 0$ and $y = l$.
The spatial dependence at non-resonance is given by
$E_a = -E_{a0} \cos(m_a y + \delta)/\cos \delta$ and
$B_a = E_{a0} \sin(m_a y + \delta)/\cos \delta$,
where $\tan \delta = \sin(m_a l)/(1 + \cos(m_a l))$ and $E_{a0} = g_{a\gamma\gamma} a_0 B_t$.

At resonance, $l = \pi/m_a$, the fields become $E_a = \sqrt{2} E_{a0} \sin(m_a y)/(m_a \delta_e)$ and
$B_a = -\sqrt{2} E_{a0} \cos(m_a y)/(m_a \delta_e)$,
where $\delta_e = \sqrt{2/(m_a \sigma)}$ is the electromagnetic penetration depth in the slabs, and $\sigma$ is their electrical conductivity.
Note that at resonance, the fields are significantly amplified because $m_a \delta_e \ll 1$.
Here we have neglected time dependence of these fields because the dependence has no important play in the discussion below. 

\vspace{0.1cm}

To facilitate absorption of the microwave, the two-dimensional electron layer should be slightly tilted rather than aligned perpendicular to the magnetic field $\vec{B}_t$.
This tilt ensures that the electric field of the axion-induced microwave has a component parallel to the layer.
Only the component is absorbed by 2D electrons.
In our setup, the surface of the sample is aligned along the $x$-direction, as shown in Fig.(\ref{b}).
We tentatively take the angle $\theta$ between the $y$-axis and the 2D electron layer to be $\pi/6$.

The applied magnetic field $\vec{B}_t$ thus has a perpendicular component
$B = B_t \cos(\theta = \pi/6) \approx 0.86 B_t$,
and a parallel component
$B_p = B_t \sin(\pi/6) = 0.5 B_t$.
Even though $\vec{B}_t$ is not perpendicular to the 2D layer, the quantum Hall effect is governed by the perpendicular component $B$.

\vspace{0.1cm}
Before discussing microwave absorption by 2D electrons, we examine how the electromagnetic wave is refracted inside the semiconductor due to its electric permittivity.
In GaAs (used in our analysis), the relative permittivity is approximately $\epsilon \simeq 13$, while $\epsilon_0 = 1$ in vacuum.

For an incident angle $\theta_i = \pi/2 - \theta = \pi/3$, the refraction angle $\theta_r$ satisfies
$\sqrt{\epsilon} \sin \theta_r = \sin \theta_i$,
yielding $\sin \theta_r \approx 0.24$.

The strengths of the electric and magnetic fields inside the semiconductor are given by

\begin{equation}
E'_a = \frac{2 \cos \theta_i}{\sqrt{\epsilon} \cos \theta_i + \cos \theta_r} E_a \approx 0.36 E_a, \quad
B'_a = \sqrt{\epsilon} E'_a \approx 1.3 E_a.
\end{equation}

The component of the electric field parallel to the 2D electron plane is
$E'_p = E'_a \cos \theta_r \approx 0.97 E'_a$.
The magnetic field $\vec{B}'_a$ inside the semiconductor is fully parallel to the 2D layer.
It is important to emphasize that the microwave field $E_a \propto g_{a\gamma\gamma} a_0 B_t$ is the value generated in vacuum, not within the semiconductor.

\vspace{0.1cm}
The component of the energy flux perpendicular to the 2D electrons is
$
E'_p B'_a \approx 0.46 E_a^2,
$ which is the portion absorbed by the 2D electron system in the quantum Hall state.

\vspace{0.1cm}
For reference, the relevant parameters are summarized as follows

\begin{equation}
E'_p \approx 0.35 E_a, \quad B_a'\approx 1.3E_a, \quad 
B \approx 0.86 B_t, \quad 
E'_p B'_a \approx 0.46 E_a^2.
\end{equation}

The cyclotron frequency is $\omega_c = eB/m^*$, the magnetic length is $l_B = \sqrt{1/eB}$, while the Zeeman energy is $g \mu_B B_t$.

\vspace{0.1cm}
Next, we analyze the absorption of axion-induced microwaves by the 2D electrons in the quantum Hall state.
A state in Landau level $n$ is described by the wave function
$
\Phi_\alpha(x,y) = e^{-iE_\alpha t} \int dk \, f_\alpha(k) \Phi(\vec{x}, n, k),
$
where $f_\alpha(k)$ is chosen such that $\Phi_\alpha$ is an eigenstate of the electron Hamiltonian including the disorder potential $V$.
The normalization condition is
$
1 = \int dx\,dy\, \overline{\Phi_\alpha(x,y)} \Phi_\alpha(x,y) = \int dk\, \overline{f_\alpha(k)} f_\alpha(k).
$

We neglect mixing between different Landau levels and between different spin states.

\vspace{0.1cm}
Since the extension of the semiconductor sample is much smaller than $1/m_a \sim 2 \text{ cm} (10^{-5} \text{ eV}/m_a)$,
we can ignore the $y$-dependent terms in $E_a$ and $B_a$, and approximate $E_a \approx E_{a0} = g_{a\gamma\gamma} a_0 B_t$.
This simplification does not affect the final results.

\vspace{0.1cm}
The absorption of the microwaves $\vec{A}'_a\sim \vec{A}'_0\exp(-im_at)$ 
with $|\vec{A}'_0|=E'_p/m_a\simeq 0.35 g_{a\gamma\gamma}a_0B_t/m_a$,
is described by
the interaction Hamiltonian, 
$H_a=\frac{-ie\vec{A}'_a\cdot\vec{P}}{m^{\ast}}$
with $\vec{P}=-i\vec{\partial}+e\vec{A}_{ex}$ and $\vec{B}=\vec{\partial}\times \vec{A}_{ex}$.

The matrix element between states $\alpha$ and $\beta$ is
$
\langle \beta | H_a | \alpha \rangle = i(E_\beta - E_\alpha) e \vec{A}'_a \cdot \vec{L}_{\alpha\beta}, \quad \vec{L}_{\alpha\beta} \equiv \langle \beta | \vec{x} | \alpha \rangle.
$

\vspace{0.1cm}
The transition rate $\tau_t^{-1}$ of an electron per unit time is

\begin{equation}
\tau_t^{-1} = 2\pi S \int dE_\beta\, \rho(E_\beta) |\langle \beta | H_a | \alpha \rangle|^2 \delta(E_\alpha + m_a - E_\beta) 
\approx 2\pi S \rho(E_\alpha + m_a) m_a^2 (e \vec{A}'_a \cdot \vec{L}_{\alpha\beta})^2.
\end{equation}

Assuming $|\vec{A}'_a \cdot \vec{L}_{\alpha\beta}|^2 \approx A_0'^2 A^2 l_B^2$, we write

\begin{equation}
\tau_t^{-1} \sim 2\pi S \rho(E_\alpha + m_a) m_a^2 e^2 A_0'^2 A^2 l_B^2.
\end{equation}
with surface area $S$ of the semiconductor sample.
The density of states is modeled, for example, as

\begin{equation}
\rho(E) = \rho_0 \exp\left(-\left( \frac{E - E_c}{\Delta E} \right)^2\right), \quad 
\rho_0 = \frac{eB}{2\pi} \cdot \frac{2}{\sqrt{\pi} \Delta E},
\end{equation}
with $E_c=\omega_c/2$, where $\int dE \rho(E)/2=eB/2\pi$ represents the number density of electrons in a Landau level.

The width $\Delta E$ is of the order of $\sim 10^{-3}$eV.
It can be estimated using the approximate formula
$
\Delta E = \sqrt{\frac{2 \omega_c}{\pi \tau_r}},
$
where $\tau_r$ denotes the relaxation time of electrons.
This relaxation time can be expressed in terms of the mobility $\mu$ of the semiconductor as
$
\tau_r = \mu m_e^{\ast}/e. 
$
Typically, $\tau_r\simeq 3.8$ps for $\mu=10^5\rm cm^2/Vs$. 
Substituting this into the expression for $\Delta E$, we find
$
\Delta E \simeq 1.4 \times 10^{-3}\text{eV} \sqrt{\left( B/10^5\text{Gauss} \right)\left( 10^5\text{cm}^2(\text{Vs})^{-1}/\mu \right)}.
$

We consider the transition from a localized state with energy $E_{\alpha}$ to a localized state
with energy $E_{\beta}=E_{\alpha}+m_a$.  Thus, $E_f<E_{\beta}=E_{\alpha}+m_a<E_c-\delta $.
The transition occurs on a plateau in $\sigma_{xy}-B$ plane, in which quantum Hall state is in an insulator phase, $\sigma_{xx}=0$.
Microwave reflection does not arise. 

\vspace{0.1cm}
The maximal number of the electrons able to absorb axion induced microwave is 

\begin{equation}
\Delta N=S\int^{E_f}_{E_f-m_a} dE \rho(E) \simeq S \rho_0 m_a \exp\left( -\left( \frac{E_f - E_c}{\Delta E} \right)^2 \right),
\end{equation}

leading to the maximal number of electrons absorbing microwaves per unit time

\begin{equation}
\Delta N \tau_t^{-1} \sim 2\pi S^2 \rho_0^2 m_a^3 e^2 A_0'^2 A^2 l_B^2 \exp\left( -2 \left( \frac{E_f - E_c}{\Delta E} \right)^2 \right).
\end{equation}
We have used $m_a=10^{-5}$eV$\sim 10^{-4}$eV $\ll E_f\sim E_c\sim10^{-2}$eV.

We compared it to the number of axion-induced photons incoming per unit time,
$
0.46 S E_a^2 / m_a \approx 3.8 S m_a A_0'^2.
$
We find that full absorption occurs when

\begin{equation}
\label{14}
\frac{\Delta N \tau_t^{-1}}{3.8 S m_a A_0'^2} \simeq 6.7\times 10^3 \Big(\frac{S}{\rm (0.1cm)^2}\Big)\Big(\frac{B}{10^5\rm Gauss}\Big)\Big(\frac{m_a}{10^{-5}\rm eV}\Big)^2\Big(\frac{10^{-3}\rm eV}{\Delta E}\Big)^2A^2 
\exp\Big(-2\Big(\frac{E_f-E_c}{\Delta E}\Big)^2\Big)> 1.
\end{equation}
The condition can be satisfied unless the overlapping factor $A$ ( $ <\beta|x|\alpha>=Al_B$ ) is not too small, e.g. $A\ll1 $.

Namely, all photons $3.8 S m_a A_0'^2$ incoming to the semiconductor sample are absorbed within a time $\tau_t$, because there are sufficiently number of electrons
$\Delta N$ which are able to absorb the photons within the time $\tau_t$. The condition in eq(\ref{14})
holds even for external microwave irradiation with frequency $m_a$ and any power.
Then, it apparently seems that incoming microwaves with any power are fully absorbed.

\vspace{0.1cm}
%Therefore, when the condition is satisfied, all of the microwaves entering the sample are absorbed.
However, for full absorption of microwaves, the incoming microwave power, which is proportional to $E^2$ (where $E$ is the electric field), must be not too large.
If the microwave power becomes high enough that the transition time $\tau_t \propto 1/E^2$ becomes shorter than the relaxation time $\tau_r$ of the 2D electrons, then the microwaves may not be fully absorbed.

\vspace{0.1cm}
The above calculation is based on the following assumption: each electron can transition independently of the transitions of other electrons. This is possible only if each electron quickly returns to its original state after the transition.
That is, it is possible when $\tau_t \gg \tau_r$.
A electron that has been excited to a higher-energy state can relax and thermalize within time $\tau_r$.
This allows the processes by which electrons in different states absorb microwaves to proceed independently. Localized electrons with energy $E_\alpha < E_f$ can absorb microwaves and transition to localized states with $E_\beta=E_{\alpha}+m_a > E_f$.
The process independently arises for each electron in energy $E_{\alpha}$ ( $E_f-m_a<E_{\alpha}<E_f$ ) .

In contrast, when $\tau_t \ll \tau_r$, the excited electrons do not have time to thermalize, and thus do not return to their original localized states within time $\tau_t$.
As a result, when two electrons in different states absorb microwaves of the same energy and undergo transitions, the processes are no longer independent. This is because the final states may not be available: once an electron has transitioned, it does not immediately return to its original state, potentially blocking subsequent transitions.	
Therefore, all electrons with energy levels in the range $E_f-m_a<E_{\alpha}<E_f$
cannot necessarily  absorb microwaves and transition to higher energy states.

%those states with energy $E_{\alpha}$ ( $E_f-m_a<E_{\alpha}<E_f$ ) remain vacant so that electrons
%with energies lower than $E_f-m_a$ should absorb the microwaves. But the number of such electrons with lower energies
%are much less than $\Delta N$. Eventually,
%there are no electrons available to absorb additional microwaves. 
%( Some of the incoming microwaves may be absorbed by excited electrons with energies above the Fermi energy $E_f$ which have already absorbed previous microwave photons. However, it remains unclear whether these excited electrons can fully absorb the incoming microwave energy. )

In particular, in cases where high-power microwaves are incident on the 2D electron system, the transition time 
$\tau_t \propto E^{-2}$ 
becomes shorter than the relaxation time $\tau_r$. 
A portion of the radiation may not be absorbed and can transmit through the sample. 

\vspace{0.1cm}

For the axion induced microwave, 
we can show that they are fully absorbed because the transition time $\tau_t $ is much longer than the
relation time even if the power of the microwave is amplified with resonance.

\begin{equation}
\tau_t \sim 10^{-1} \text{ s} \left( \frac{\text{(0.1cm)}^2}{S} \right) \left( \frac{m_a}{10^{-5} \text{ eV}} \right) \left( \frac{10^5 \text{ Gauss}}{B} \right)^2 \left( \frac{1}{A^2} \right)\left(\frac{0.3\,\text{GeV}/\text{cm}^3}{\rho_d}\right)
\exp\left( \left( \frac{E_f - E_c}{\Delta E} \right)^2 \right),
\end{equation}
which is much longer than relaxation time of the order of $\tau_r \sim 10^{-12} \text{ s}$, unless $A$ is much large $\gg 1$.

Hence, axion-induced microwaves are fully absorbed by the 2D electron gas in the quantum Hall regime.
As a result, the sample's temperature increases, albeit only slightly in the absence of resonance.
The incoming power is tiny

\begin{equation}
P_{ra} \sim 0.46 S E_a^2 \sim 10^{-32} \text{ W} 
\left( \frac{B}{10^5 \text{ Gauss}} \right)^2 \left( \frac{S}{ \text{(0.1cm)}^2} \right)\left(\frac{\rho_d}{0.3\,\text{GeV}/\text{cm}^3}\right),
\end{equation}

but it can be amplified to $\sim 10^{-20} \text{ W}$ using a properly designed resonant cavity.

\vspace{0.1cm}

\section{temperature increase of sample}

We show that the absorption of the power $P_{ra}$, estimated above, slightly increases the temperature of the sample exhibiting the quantum Hall effect. However, this power is not large enough to produce a detectable temperature rise.

Now, we consider the effect of resonance. We demonstrate that the temperature increase becomes sufficiently large to be detectable when resonance occurs in the cavity, achieved by tuning an appropriate length scale.

In the case of two parallel slabs, ( see Fig.(\ref{b}) ), resonance takes place when the separation between the slabs is
$l=\frac{\pi}{m_a} \simeq 6.3\,\mbox{cm} \big( \frac{10^{-5}\,\rm{eV}}{m_a} \big)$.

\vspace{0.1cm}

When resonance occurs, the axion-induced microwave field in vacuum (with $B_a \sim E_a$) is amplified \cite{iwazaki01} as

\begin{equation}
E_a \to \frac{E_a}{m_a \delta_e}, \quad \text{with penetration depth} \quad \delta_e = \sqrt{\frac{2}{m_a \sigma}},
\end{equation}
where $\sigma$ is the electrical conductivity of the metal forming the cavity (the slabs). For example, for 6N copper at temperatures below 1 K, $\sigma \simeq 3.3 \times 10^7\,\text{eV}$ (corresponding to $\sigma^{-1} = 0.2 \times 10^{-11}\,\Omega\cdot\text{m}$).

This leads to the enhancement factor

\begin{equation}
\frac{1}{m_a \delta_e} \simeq 1.3 \times 10^6 \sqrt{ \left( \frac{\sigma}{3.3 \times 10^7\,\text{eV}} \right) \left( \frac{10^{-5}\,\text{eV}}{m_a} \right) }.
\end{equation}

The electric field $E'_p$ in the semiconductor, as well as the magnetic field $B'_a$, is enhanced by the same factor $1/(m_a \delta_e)$.

\vspace{0.1cm}

As a result, the amplified energy flux entering the sample is

\begin{equation}
\frac{E'_p B'_a}{(m_a \delta_e)^2} \simeq 0.46 \left( \frac{E_a}{m_a \delta_e} \right)^2,
\end{equation}

and it is absorbed in the quantum Hall state. The power absorbed by the sample with surface area $S$ is given by

\begin{equation}
P_{ra} \sim 0.46 \left( \frac{E_a}{m_a \delta_e} \right)^2 S \sim 0.9 \times 10^{-20}\,\text{W}\,g_\gamma^2 
\Big(\frac{S}{\text{(0.1cm)}^2}\Big)
\left( \frac{10^{-5}\,\text{eV}}{m_a} \right)
\left( \frac{\sigma}{10^7\,\text{eV}} \right)
\left( \frac{B_t}{1.5 \times 10^5\,\text{Gauss}} \right)^2
\left( \frac{\rho_d}{0.3\,\text{GeV/cm}^3} \right),
\end{equation}
with
$E_a^2 = (g_{a\gamma\gamma} a_0 B_t)^2 \simeq 4.0 \times 10^{-30}g_{\gamma}^2
\,\text{W/cm}^2 \left( \frac{B_t}{1.5 \times 10^5\,\text{Gauss}} \right)^2$.

\vspace{0.1cm}
It should be contrasted with the power absorbed in conductor with large electrical conductivity. 
Electric field in the conductor is suppressed by factor $m_a\delta_e$\cite{iwazaki01, kishimoto,iwazaki02} compared with the one in vacuum.
Thus, the power is enhanced only by the factor $1/m_a\delta_e\sim 10^6$ when the resonance occurs. The suppression simply arises
owing to the reflection of incident radiations by the conductor. Almost of all the incident radiations with large amplitude $\sim 10^{12}$
are reflected
so that only small fraction $\sim 10^6$ penetrates in the metal.
\vspace{0.1cm}

To confirm the enhancement factor $\frac{1}{(m_a \delta_e)^2}\sim 10^{12}$,
we have performed calculations that take into account both the velocity distribution of axions and the frequency spread of the resonance peak associated with the quality factor of the resonant cavity. The velocity distribution is characterized by a quality factor ($Q_a \sim 10^6$), while the resonance of the cavity has a width characterized by ($Q_l \sim 10^6$). As a result, we have confirmed that a small semiconductor sample placed inside the cavity experiences a large enhancement of the electromagnetic flux, as described in the paper, of order ($Q_a Q_l \sim 10^{12}$). It is obvious to have such a result because semiconductor at low temperature $\sim 10$mK is
an insulator just like a vacuum. Almost of all the radiations can penetrate the semiconductor.
Two dimensional electrons absorb the amplified radiations without depressing the magnitude of the 
radiations inside the thin quantum well $\sim 0.01\mu$m trapping the electrons. On the other hand, the radiations are depressed inside
skin width $\delta \sim 0.01\mu$m in the metal; the radiations disappear from the surface to deeper layers of the metal. 
That is, in the case of the metal, most of the radiations are reflected by the metal. On the other hand, two dimensional electrons do not
make the radiation reflect because electrons are only present in the quantum well. As we have explained, 
whole radiations generated by axion are absorbed by two dimensional electrons forming quantum Hall state, as far as
the number of incident photons is not so large that electrons cannot absorb them.

\vspace{0.1cm}
The temperature of the sample increases due to this absorbed energy.
To estimate the temperature rise, we need to know the heat capacities of both the semiconductor and the 2D electrons.
According to Debye model, 
the heat capacity $C_s$ of a rectangular semiconductor with volume $Sd$ at low temperatures (below 100 mK) is

\begin{equation}
C_s \simeq 1.94 \times 10^3\,\text{J/g·K} \left( \frac{T}{T_D} \right)^3 \left( \frac{\text{density}}{M} \right) Sd,
\end{equation}
where $M$ is the molecular weight of the semiconductor and $T_D$ is the Debye temperature. For GaAs, $M = 144$, density = $5.3\,\text{g/cm}^3$, and $T_D = 360\,\text{K}$.
(We assume that the semiconductor's heat capacity is dominated by phonons.)

Numerically, this gives

\begin{equation}
C_s \simeq 0.88 \times 10^6 \left( \frac{T}{20\,\text{mK}} \right)^3 \left( \frac{S}{\text{(0.1cm)}^2} \right) \left( \frac{d}{1\,\mu\text{m}} \right).
\end{equation}

On the other hand, the heat capacity of 2D electrons in the low-temperature limit is

\begin{equation}
C_e = \frac{dQ}{dT} \simeq \frac{\pi^2}{3} T S \rho(E_f) = \frac{\pi^2}{3} T S \rho_0 \exp\left( -\left( \frac{E_f - E_c}{\Delta E} \right)^2 \right),
\end{equation}

where $Q$ is the energy of the 2D electron system,

\begin{equation}
Q = \int_0^{\infty} dE\, E\, \rho(E) \frac{1}{\exp\left(\frac{E - E_f}{T}\right) + 1}.
\end{equation}

Numerically, this yields

\begin{equation}
C_e \simeq 3.4 \times 10^6 \left( \frac{S}{\text{(0.1cm)}^2} \right) \left( \frac{T}{20\,\text{mK}} \right) \left( \frac{B_t}{1.5 \times 10^5\,\text{Gauss}} \right)^{1/2}
\exp\left( -\left( \frac{E_f - E_c}{\Delta E} \right)^2 \right).
\end{equation}

Thus, when $\exp\left( - (E_f - E_c)^2 / \Delta E^2 \right) < 10^{-1}$, the dominant contribution to the total heat capacity is from the semiconductor, i.e., $C_s + C_e \approx C_s$.

However, note the condition in Eq.(\ref{14}), which requires 
$\Big(\frac{10^{-3}\rm eV}{\Delta E}\Big)A\exp\left( - (E_f - E_c)^2 / \Delta E^2 \right)$ not to be too small. This condition must be satisfied for the axion-induced microwave to be completely absorbed.

Therefore, it is likely that choosing the Fermi energy $E_f$ such that 

\begin{equation}
10^{-3}\Big(\frac{10^{-3}\rm eV}{\Delta E}\Big)^{-1}A^{-1} < \exp\left( -\left( \frac{E_f - E_c}{\Delta E} \right)^2 \right) < 10^{-1}
\end{equation}
which ensures both that $C_s > C_e$ and that the microwave is effectively absorbed.
It is favorable to have smaller width $\Delta E <10^{-3}$eV, which is realized with larger mobility $\mu>10^5\rm cm^2/Vs$. 
The Fermi energy
is approximately in the range
$E_f \simeq E_c - (1.5 \sim 2.6)\Delta E$ with $\Big(\frac{10^{-3}\rm eV}{\Delta E}\Big)A=1$, 
that is, in the tail of the density of states $\rho(E_f)$.

\vspace{0.1cm}
In actual experiments, we control not the Fermi energy $E_f$ directly, but the external magnetic field $B_t$. To meet the conditions described above, we must choose the magnetic field such that the resulting Fermi energy falls within the required range.

For example, with $B_t = 1.5 \times 10^5\,\text{Gauss}$, we find $E_c = \omega_c / 2 \simeq 2.3 \times 10^{-2}\,\text{eV}$, and the Zeeman energy is about $3.9 \times 10^{-4}\,\text{eV}$. Since the Zeeman splitting is much smaller than the cyclotron energy $\omega_c$, the quantum Hall plateaus at $\nu = 1$ and $\nu = 2$ are nearly degenerate. We must appropriately choose the magnetic field so that $E_f$ is near the edge (i.e., $\nu < 1$) of the $\nu = 1 \sim 2$ plateau in $\rho_{xy}$ or $\sigma_{xy}$. 

\vspace{0.1cm}
In order to find appropriate magnetic field, 
we test the temperature response of the sample under weak external microwave irradiation with power on the order of $\sim 10^{-18}\,\text{W}$. By appropriately tuning the magnetic field $B_t$, full absorption arises, resulting in a detectable temperature rise. This preparatory step helps us optimize the strength of the magnetic field, in other words, Fermi energy $E_f$.

\vspace{0.1cm}

Now, when the sample absorbs the amplified microwave radiation, the rate of temperature increase is given by

\begin{equation}
\dot{T} = \frac{P_{ra}}{C_s(T)} \simeq \frac{0.72\,\text{mK}}{\text{s}}\,g_{\gamma}^2\left(\frac{20\,\text{mK}}{T}\right)^3
\left( \frac{\sigma}{10^7\,\text{eV}} \right)\left(\frac{10^{-5}\,\text{eV}}{m_a}\right)\left(\frac{1\,\mu\text{m}}{d}\right)
\left(\frac{B_t}{1.5\times10^5\,\text{Gauss}}\right)^2\left(\frac{\rho_d}{0.3\,\text{GeV}/\text{cm}^3}\right)
\end{equation}
where we take the thickness $d = 1\,\mu\text{m}$ and temperature $T = 20\,\text{mK}$, with an external magnetic field 
$B_t = 1.5 \times 10^5\,\text{Gauss}$, applied parallel to the sample layers.
( Even when smaller magnetic field $B_t=10^5$Gauss, $\dot{T}\simeq 0.32$mK/s with lower temperature $T=20$mK. )

\vspace{0.1cm}
The thermal energy generated from microwave absorption dissipates into the thermal bath. In the actual experiment, we measure the temperature increase in the presence of this thermal dissipation.

\vspace{0.1cm}

The temperature of the sample, $T + \Delta T$ (with $T = 20\,\text{mK}$), evolves according to the differential equation

\begin{equation}
C_s \frac{d\Delta T}{dt} = -G \Delta T + P_{ra}
\end{equation}
with initial condition $\Delta T(t=0)=0$,
where $G$ denotes the thermal conductance. Thermal energy of semiconductor sample leaks to heat bath with the rate of 
thermal conductance $G$.
Since $C_s \propto T^3$ and $G \propto T^3$, the stationary solution is

\begin{equation}
\Delta T(t=\infty) = \frac{P_{ra}\tau}{C_s(\Delta T(t=\infty) + 20\,\text{mK})}\simeq  \frac{P_{ra}\tau}{C_s(20\,\text{mK})}
\,\,\,\text{for}\,\,\,\Delta T\ll 20\text{mK}
\end{equation}
where $\tau \equiv C_s/G$ is the time constant such that without microwave power ($P_{ra} = 0$), the excess temperature decays as 
$\Delta T \propto \exp(-t/\tau)$ as $t \to \infty$.

\vspace{0.1cm}
We should make a comment that actual thermal conductance $G$ can be made much small. For example, the sample is connected to 
heat bath with superconducting nanowire lead and thin film pedestal supporting the sample in dilution refrigerator. 
Superconducting nanowires, e.g. NbTi, are used in low temperature physics. 
For instance, their radii ( length ) are of the order of $1\mu$m ( $1$cm ) and thermal conductance of NbTi, e.g. $G\sim 0.01\text{W/Km}$
at $T=4$K. 
Then we can have $G=0.01\text{W/Km}(20\text{mK}/4\text{K})^3\pi(1\mu m)2/1\text{cm} 
\sim 10^{-18}\text{W/K}$ at $T=20$mK so that $\tau=C_s/G\sim 10$s. 
Similarly, when we use thin film pedestal of SiN with thickness $100$nm, width $1\mu$m and length $100\mu$m,
the thermal conductance is of the same order of the one as superconducting nanowire lead.
The smallness of the thermal conductance $G$ arises from smallness of the wire lead and pedestal, and
more importantly the low temperature $T=20$mK, because 
thermal energies are carried by phonons, not electrons so that $G\propto T^3$.
Therefore, the time constant  $\tau=C_s/G$ can be made larger than $1$ second. We measure
the temperature rise $P_{ra}t_{ob}/G$ with observational time $t_{ob}$ smaller than the time constant $\tau >1\text{s}$.

%\begin{equation}
%\Delta T = \frac{P_{ra}}{C_s(T)} \simeq 0.72\,\text{mK}\Big(\frac{t_{ob}}{\text{s}}\Big)\,g_{\gamma}^2\left(\frac{20\,\text{mK}}{T}\right)^3
%\left( \frac{\sigma}{10^7\,\text{eV}} \right)\left(\frac{10^{-5}\,\text{eV}}{m_a}\right)\left(\frac{1\,\mu\text{m}}{d}\right)
%\left(\frac{B_t}{1.5\times10^5\,\text{Gauss}}\right)^2\left(\frac{\rho_d}{0.3\,\text{GeV}/\text{cm}^3}\right)
%\end{equation}
%  

\vspace{0.1cm}
In order to measure the temperature rise, it is favorable to use a thermometer\cite{qpc} based on a quantum point contact (QPC), which can directly measure the temperature of two-dimensional electrons. The thermal conductance associated with the thermometer 
is much small similar to the one of superconducting nanowire. Thus, the thermometer does not almost leak heat energy from 2D electrons. 
In principle, we can measure the temperature rise $\Delta T$ within accuracy less than $0.01$mK with the thermometer. 
The realistic accuracy, e.g. $0.3$mK determines the excluded region in the plane $g_{a\gamma\gamma}-m_a$, as explained below.

\vspace{0.1cm}
Thus, the temperature rise we observe with the observation time $t_{ob}(<\tau)$ is approximately

\begin{eqnarray}
\label{30}
\Delta T(t=\infty)
\simeq &0.72&\,\text{mK}\left(\frac{t_{ob}}{1\,\text{s}}\right)\,g_{\gamma}^2\left(\frac{10^{-5}\,\text{eV}}{m_a}\right)
\left(\frac{20\text{mK}}{T}\right)^3\left( \frac{\sigma}{10^7\,\text{eV}} \right) \nonumber \\
&\times& \left(\frac{1\,\mu\text{m}}{d}\right)
\left(\frac{B_t}{1.5\times10^5\,\text{Gauss}}\right)^2\left(\frac{\rho_d}{0.3\,\text{GeV}\rm cm^{-3}}\right) .
\end{eqnarray}

The formula is applicable only for QCD axion model in which the coupling constant $g_{a\gamma\gamma}$ satisfies
$g_{a\gamma\gamma}\simeq 3.9g_{\gamma}(10^{-15}\text{GeV}^{-1})(m_a/10^{-5}\text{eV})$. Using model independent coupling 
constant $g_{a\gamma\gamma}$, we represent the temperature increase in the following,

\begin{eqnarray}
\Delta T(t=t_{ob})
\simeq &0.05&\text{mK}\left(\frac{t_{ob}}{1\,\text{s}}\right)\,\Big(\frac{g_{a\gamma\gamma}}{10^{-15}\text{GeV}^{-1}}\Big)^2
\left(\frac{20\text{mK}}{T}\right)^3\left(\frac{10^{-5}\,\text{eV}}{m_a}\right)^3\left( \frac{\sigma}{10^7\,\text{eV}} \right) 
\nonumber \\
&\times&\left(\frac{1\,\mu\text{m}}{d}\right)
\left(\frac{B_t}{1.5\times10^5\,\text{Gauss}}\right)^2\left(\frac{\rho_d}{0.3\,\text{GeV}\rm cm^{-3}}\right) .
\end{eqnarray}

\begin{figure}[htp]
\centering
\includegraphics[width=0.6\hsize]{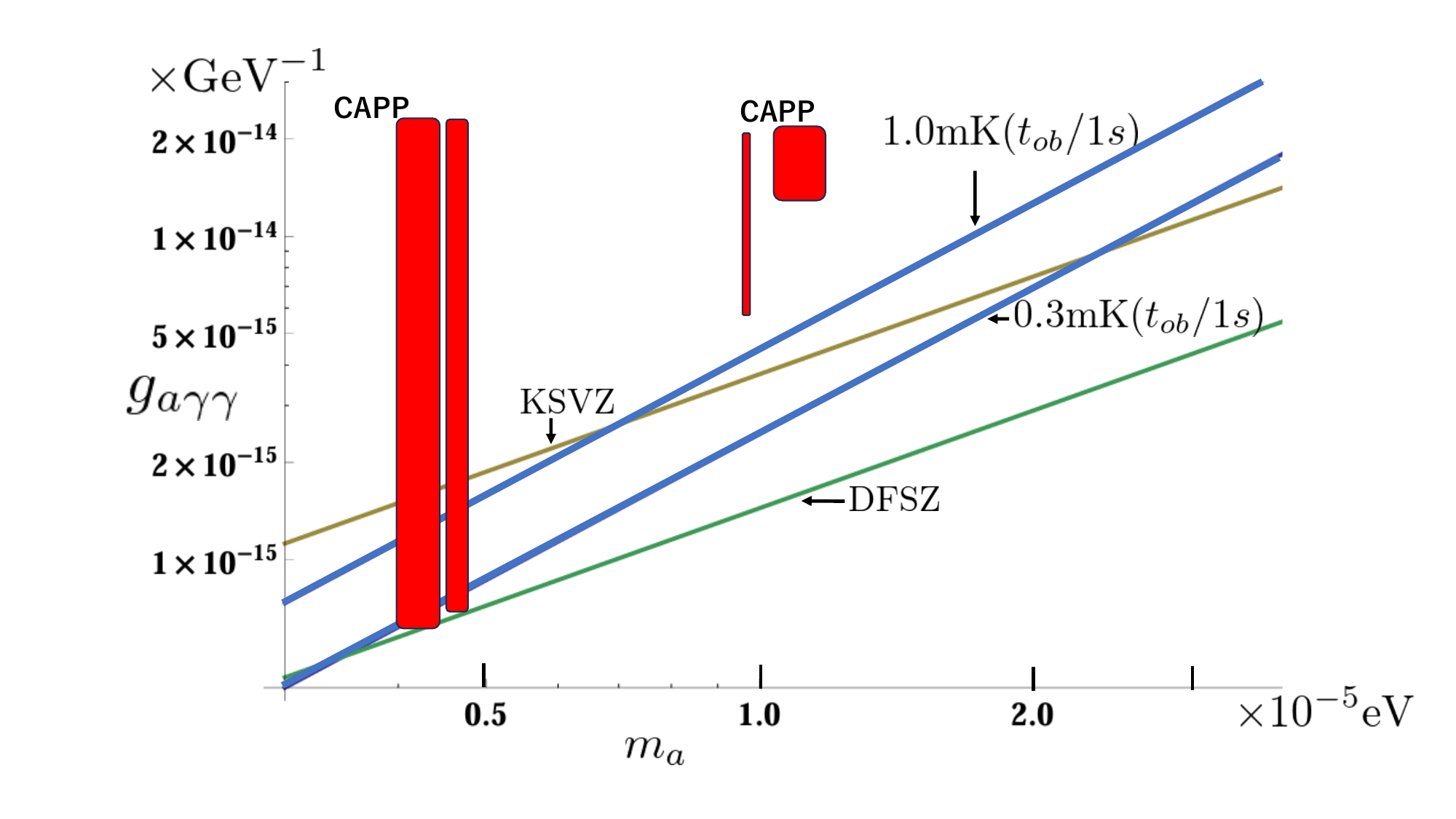}
\caption{When we do not observe $\Delta T$ with accuracy $0.3\text{mK}(t_{ob}/1\text{s})$ or $1.0\text{mK}(t_{ob}/1\text{s})$ with observation time $t_{ob}=1$s, the value of $g_{a\gamma\gamma}$ above the thick lines is excluded. We roughly show excluded region obtained by CAPP.}
\label{c}
\end{figure}

We show in Fig.(\ref{c}) the excluded region of $g_{a\gamma\gamma}$ and $m_a$ indicated by thick lines
$0.3\text{mK}(t_{ob}/1\text{s})$ and $1.0\text{mK}(t_{ob}/1\text{s})$ with observation time $t_{ob}=1$s. The lines
show the accuracy of the measurement.
%If the time constant $\tau$ is less than $10$ms, for instance, $1$ms, the excluded region is above the same
%dashed line indicated with $\Delta T= 0.05\text{mK}(\tau/1\text{ms})$ as the line in the figure 
%when we do not observe $\Delta T\ge 0.05$mK.   
In the figure, we have roughly depicted the excluded region by CAPP\cite{CAPP1,CAPP2,CAPP3}.
We also show the lines associated with the models of QCD axion. 

\vspace{0.1cm}
We consider measuring the temperature change while scanning the frequency in $1$-second intervals. 
When we do not detect temperature rise within the accuracy, for example, $0.3$mK at a frequency, 
the region above the line denoted with $0.3\text{mK}(t_{ob}/1s)$ at the frequency is excluded.
Improving the accuracy such as $0.1\text{mK}(t_{ob}/1s)$, the excluded region is expanded much below the 
line  $0.3\text{mK}(t_{ob}/1s)$. On the other hand,
If a temperature increase of $0.5$ mK within one second is observed at a particular frequency, 
the axion mass can be inferred from that frequency, and the coupling 
$g_{a\gamma\gamma}$ can be determined from the magnitude of the temperature increase.

Therefore, the higher the accuracy of the temperature measurement, the wider the area in $g_{a\gamma\gamma}-m_a$ plane 
that can be investigated.
For the purpose, a large time constant $\tau > 1\,\text{s}$ is preferable 
to observe a measurable temperature increase $\Delta T$. 
To achieve such a large time constant, it is needed 
to minimize thermal conductance $G$ using possibly small superconducting nanowire and thin film pedestal.
Furthermore, it is preferable to use a thermometer\cite{qpc} based on a quantum point contact (QPC) with possibly fine accuracy
to search large region in the plane $g_{a\gamma\gamma}-m_a$.

\vspace{0.1cm}

We have focused on axion detection around a mass of 
$10^{-5}$eV. 
However, as indicated by these equations, 
axions with smaller masses are expected to produce larger temperature increases and are therefore easier to detect.
In particular, we have large temperature increase $\Delta T\sim 50$mK when $m_a=10^{-6}$eV.

\vspace{0.1cm}

To identify the resonance condition, we finely tune the cavity length $l$. 
Since the axion energy $m_a + \frac{1}{2}m_a v^2$ (with $v \sim 10^{-3}$) is distributed around $m_a$ 
with a relative width of $\sim 10^{-6}$, the cavity length must be scanned in steps of
$\delta l= 10^{-6} ( \pi/m_a ) \simeq 6.3 \times 10^{-6}\,\text{cm} \left(10^{-5}\,\text{eV}/m_a\right)$.

If we measure the sample temperature for 1 second at each step, scanning the axion mass range $(0.9 \sim 1.1) \times 10^{-5}\,\text{eV}$ takes approximately 50 hours. While the measurement time per step may vary depending on the detection method, this approach provides a highly efficient means of searching for axions.

%\vspace{0.1cm} 
%To detect axion,
%we measure the increase of the temperature of material caused by the absorption of amplified but still much weak microwaves generated by the axion. So it is favorable to use material with much small heat capacity and small thermal conductance G
%of the material connecting to heat bath.
%In the paper we have proposed two dimensional electrons forming quantum Hall states as such a material.
%But the idea is also applicable for materials which are able to absorb the microwaves and have sufficiently small heat capacities
%for the temperature to increase. 
%To find such a material is a subject in near future.
% 

\section{conclusion}

We have proposed a novel method for axion detection that leverages the quantum Hall effect within an axion haloscope. By precisely tuning the haloscope's dimensions, axion-induced microwave radiation can be resonantly amplified. This amplified radiation is absorbed by a semiconductor sample exhibiting the quantum Hall effect, leading to a measurable increase in temperature—a signal that would otherwise be undetectable without amplification.

Using a GaAs-based sample with small thickness $d=1\mu$m, 
we have demonstrated that the temperature rise can reach approximately
$0.05\mathrm{mK}\big(t_{ob}/1\mathrm{s}\big)\big(1\mu \mbox{m}/d\big)(g_{a\gamma\gamma}/10^{-15}\text{GeV}^{-1})^2
(10^{-5}\,\text{eV}/m_a)^3(20\text{mK}/T)^3( \sigma/10^7\,\text{eV}) $
at a base temperature of $20$mK under an external magnetic field of $1.5\times 10^5$Gauss, where the observation time $t_{ob}$ is 
supported to shorter than the thermal relaxation time $\tau$ of the energy dissipation into heat bath. 
We need longer relaxation time $\tau$ for observation time $t_{ob}$ taken to be longer. Such a large $\tau$ can be realized
by using superconducting nanowire lead and thin film pedestal supporting the sample in dilution refrigerator.   
The detection of small temperature increase may be possible by using QPC thermometer.
%The estimation is obtained using our proposed axion haloscope configuration,
%consisting of parallel slabs
%with an electric conductivity $\sigma \simeq 3.3 \times 10^7\,\text{eV}$. Similar results would be
%obtained with other axion haloscopes even if the quality factor $Q$ value in the haloscope is small. 
The method proposed here makes it easier to detect axions with a mass $\sim 10^{-5}$ eV, which are difficult to detect with conventional 
halo scopes.

\vspace{0.2cm}
The author
expresses thanks to A. Sawada for useful comments.

\end{document}